\documentclass[aps,prd,amsmath,amssymb,footinbib]{revtex4-1}

\usepackage{graphicx}
\usepackage{natbib}

\usepackage{bm}
\usepackage{url}
\usepackage{textcase}
\usepackage{subcaption}
\usepackage{float}
\usepackage{amsthm}
\usepackage{braket}
\usepackage{color}

\begin{document}

\title{Rotating tensor-multiscalar black holes with two scalars}

\author{Lucas G. Collodel}
 \email{lucas.gardai-collodel@uni-tuebingen.de}

\affiliation{Theoretical Astrophysics, Eberhard Karls University of T\"ubingen, T\"ubingen 72076, Germany}

\author{Daniela D. Doneva}
\email{daniela.doneva@uni-tuebingen.de}
\affiliation{Theoretical Astrophysics, Eberhard Karls University of T\"ubingen, T\"ubingen 72076, Germany}
\affiliation{INRNE - Bulgarian Academy of Sciences, 1784  Sofia, Bulgaria}

\author{Stoytcho S. Yazadjiev}
\email{yazad@phys.uni-sofia.bg}
\affiliation{Theoretical Astrophysics, Eberhard Karls University of T\"ubingen, T\"ubingen 72076, Germany}
\affiliation{Department of Theoretical Physics, Faculty of Physics, Sofia University, Sofia 1164, Bulgaria}
\affiliation{Institute of Mathematics and Informatics, 	Bulgarian Academy of Sciences, 	Acad. G. Bonchev St. 8, Sofia 1113, Bulgaria}

\begin{abstract}
We construct hairy black hole solutions in a particular set of tensor-multi-scalar theories of gravity for which the target-space admits a Killing vector field with periodic flow that is furthermore the generator of a one-parameter family of point transformations which leave the whole theory invariant. As for black holes with scalar hair in general relativity, these must be rotating objects as the scalarization arises from superradiant instability. We analyze the most fundamental rotating solutions (winding number $m=1$) in five theories which differ from each other by the value of the Gauss curvature of the target-space. We investigate the domain of existence of these solutions, i.e. how they depart from Kerr black holes or solitons, and verify that there is a non-uniqueness region where for a fixed pair of mass and angular momentum, different black holes can be found. Based on the ration between the mass stored in the scalar hair and the horizon mass we access how hairy this solutions are and show the with increasing Gauss curvature, hairier solutions are found. Interestingly, the smaller this curvature is, the higher is the minimum ratio $J/M^2$ featured by the solution set. The horizon area is calculated and in the non-uniqueness domain some (but not all) hairy black holes are more thermodynamically stable than bald Kerr black holes. Finally, we analyze how bulgy the horizon of these hairy black holes are through the ratio between the equatorial and polar circumferences and we observe that for none of these solutions it overcomes the deformation of extremal Kerr black holes.
\end{abstract}

\maketitle

\section{INTRODUCTION}
The newly emerging field of gravitational wave astronomy is gathering more and more momentum and it is offering completely new possibilities to probe the nature of black holes and the strong field regime of gravity in general \cite{TheLIGOScientific:2016src,Isi:2019aib}. A careful and high precision observation of the gravitational wave signal, including higher order modes, can tell apart the Kerr black hole from its generalizations \cite{Brito:2018rfr,Carullo:2018sfu,Forteza:2020hbw,Cook:2020otn,Giesler:2019uxc,Volkel:2019muj,Volkel:2020daa}. It is not trivial, though, to identify the theories that allow for the existence of non-Kerr black holes and to actually construct such solutions analytically or numerically. The reason is that for a very large class of alternative theories of gravity the so-called no-hair theorems exist \cite{Heusler:1996ft,Robinson:2004zz} which state that a black hole should be completely characterized but its mass, angular momentum and charge. There are different possibilities to evade the no-hair theorem \cite{Herdeiro:2015waa,Sotiriou:2015pka,Cardoso:2016ryw} but a careful analysis shows that in many cases the obtained black hole solutions are unstable, non-asymptotically flat or suffer from other pathologies. 

A very interesting, and one of the most viable possibilities for the development of scalar hair was discovered a few years ago based on the assumption that the additional scalar field do not inherit the symmetries of the spacetime. More precisely, one allows for the existence of axisymmetric time-dependent complex scalar field which has a harmonic dependence on $t$ and $\phi$. Indeed, at the threshold of supperradiance, when the frequency of the scalar field is equal to an integer times the angular velocity of the horizon, black holes with the so-called synchronized scalar hair can exist. This was demonstrated for the first time in \cite{Hod:2012px,Hod:2013zza} in the perturbative regime (see also \cite{Benone:2014ssa,Hod:2016yxg}) and later the nonlinear solutions were constructed \cite{Herdeiro:2014goa,Herdeiro_2015}. Further extensions of these results were made in a series of papers \cite{Brihaye:2014nba,Kleihaus:2015iea,Herdeiro:2015kha,Herdeiro:2015tia,Herdeiro:2016tmi,Delgado:2016jxq,Herdeiro:2018wvd,Santos:2020pmh,Kunz:2019bhm,Herdeiro:2014pka,Herdeiro:2017oyt,Kunz:2019sgn}.

A class of alternative theories of gravity that allows for the existence of such hairy black holes, but offer on the other hand much richer phenomenology, is the tensor-multi-scalar theories of gravity \cite{Damour1992,Horbatsch2015}. This is one of the few generalizations of Einstein's theory that possesses some of the most desired properties -- it is very natural from a physical point of view, the presence of multiple scalar fields is motivated by fundamental theories, like the theories trying to unify all the interactions, it does not suffer from any severe intrinsic problems like instabilities or badly posed Cauchy problem, and last but not leasts -- with a proper choice of the parameters it is in agreement with all the present observations. Even though having more than one scalar field introduces significant technical difficulties, it offers new and very rich phenomenology. Various objects have be constructed in these theories such as  solitons and mixed soliton-fermion stars \cite{Yazadjiev2019,Collodel:2019uns,Doneva:2019krb} and topological and scalarized neutron stars \cite{Doneva:2019ltb,Horbatsch2015,Doneva:2020afj}. A careful analysis of the field equations reveals the very interesting possibility for the existence of black holes with synchronized scalar hair that have as limit the rotating solitons obtained in \cite{Collodel:2019uns}. The present paper is devoted on constructing and examining the properties of these solutions. We will demonstrate that, when the target space of the multiple scalar fields is flat, these solutions reduce to the solutions obtained by Herdeiro and Radu \cite{Herdeiro:2014goa,Herdeiro_2015}, while for nonzero, and especially negative, curvature of the target space metric the black hole properties have significant qualitative differences. 

In Section II we review the basics of the particular class of tensor-multi-scalar theories of gravity we will be working with, as well as the boundary conditions and the black hole global charges. The reduced field equations describing rotating black hole solutions are given in a separate Appendix. The numerical results are presented and discussed in Section III. The paper ends with conclusions. 

Except where required for clarification, geometrized units are adopted $c=G=1$.

\section{THEORY}
\label{s1}

\subsection{Tensor-multi-scalar theories of gravity}

General TMST account for the existence of any number $N$ of scalar fields which act as extra gravitational degrees of freedom and form the basis of a coordinate patch of a target space ${\cal E}_N$ described by its own metric field $\gamma_{ab}$ of dimension $N$. 
In the Einstein frame, we can describe this class of theories via the following action,
\begin{equation}
\label{action}
S=\frac{1}{16\pi G_{\ast}}\int \left[R-2g^{\mu\nu}\gamma_{ab}(\varphi)\partial_\mu\varphi^a\partial_\nu\varphi^b-4V(\varphi)\right]\sqrt{-g}d^4x+S_M[A^2(\varphi)g_{\mu\nu};\Psi], 
\end{equation}
where $R$ and $g$ are the Ricci curvature and the metric determinant of the corresponding frame, respectively, $G_{\ast}$ is the bare gravitational constant, $\varphi^a$ is the $a$-th scalar field and $V(\varphi)$ is the potential for the scalar sector. Greek indices are used for spacetime components and latin indices for scalars. Matter fields $\Psi$ are described within the $S_M$ action, and are minimally coupled to gravity via the Jordan frame metric $\tilde{g}_{\mu\nu}=A^2(\varphi) g_{\mu\nu}$. Therefore, a particular theory is defined by specifying the number of scalar fields, the form of the target space, the potential and the conformal factor $A(\varphi)$ along with all the parameters that should appear therein. 

The field equations, in the Einstein frame, are obtained as usual by varying the action with respect to $g^{\mu\nu}$
\begin{equation}
\label{efe}
R_{\mu\nu}=2\gamma_{ab}\partial_\mu\varphi^a\partial_\nu\varphi^b+2V(\varphi)g_{\mu\nu},
\end{equation}
and a set of Klein-Gordon-Einstein (KGE) equations arises from varying it with respect to each scalar field $\varphi^a$,
\begin{equation}
\label{kge}
\Box\varphi^a=-\gamma^a_{bc}(\varphi)g^{\mu\nu}\partial_\mu\varphi^b\partial_\nu\varphi^c+\gamma^{ab}(\varphi)\frac{\partial V(\varphi)}{\partial\varphi^b},
\end{equation}
where $\gamma^a_{bc}(\varphi)$ is the Christoffel symbols for the metric $\gamma_{ab}(\varphi)$ of the target space. 

As a continuation of our previous works \cite{Yazadjiev2019,Collodel:2019uns}, in this article we focus on $N=2$ theories for which the target space is maximally symmetric and possesses a Killing field with a periodic flow, $K^a$, that acts trivially upon the potential and conformal factor, i.e. $K^a\partial_a A(\varphi)=K^a\partial_a V(\varphi)=0$. These theories are interesting for their simplicity while still allowing for the existence of hairy rotating black holes. They are  invariant under point transformations generated by $K^a$, and therefore there exists a conserved Noether current \cite{Yazadjiev2019}, given by
\begin{equation}
\label{nc}
\tilde{j}^\mu=\frac{1}{4\pi G(\varphi)}\tilde{g}^{\mu\nu}K_a\partial_\nu\varphi^a,
\end{equation} 
with $G(\varphi)=G_{*}A^2(\varphi)$.

The two dimensional metric of the target space is written in a conformally flat form, $\gamma_{ab}=\Omega^2(\varphi)\delta_{ab}$, where $\delta_{ab}$ is the Kroenecker delta, and
\begin{equation}
\Omega^2=\frac{1}{\left(1+\frac{\kappa}{4}\delta_{ab}\varphi^a\varphi^b\right)^2},
\end{equation}
where $\kappa$ is the Gauss curvature of the target manifold. In these coordinates, the Killing field is spelled out as $K=\varphi^{(2)}\frac{\partial}{\partial\varphi^{(1)}}-\varphi^{(1)}\frac{\partial}{\partial\varphi^{(2)}}$, and, in order to preserve invariance, both the potential and the conformal factor must be functions of the combination $\psi^2=\delta_{ab}\varphi^a\varphi^b$. We presently consider only massive free scalar fields, for which the potential reads
\begin{equation}
V(\psi)=\frac{1}{2}\mu^2\psi^2,
\end{equation}
and although we are investigating only vacuum solution ($S_M=0$), the conformal factor still needs to be specified for calculating physical quantities that are not invariant when transforming from the Einstein to the Jordan frame, i.e. those that depend explicitly on the metric functions $\tilde{g}_{\mu\nu}$ such as geometrical measures like lengths, areas and volumes. Conservatively, we adopt
\begin{equation}     
\label{conffactor}       
A(\psi)=\exp\left(\frac{1}{2}\beta\psi^2\right).
\end{equation}

We seek stationary and axisymmetric solutions containing a null hypersurface, and with adapted spherical coordinates $(t,r,\theta,\phi)$ we choose the following Ansatz for the line element
\begin{equation}
\label{metric}
ds^2=-\mathcal{N}e^{2F_0}dt^2+e^{2F_1}\left(\frac{dr^2}{\mathcal{N}}+r^2d\theta^2\right)+e^{2F_2}r^2\sin^2\theta\left(d\phi-\frac{\omega}{r}dt\right)^2,
\end{equation}
where $\mathcal{N}=1-r_H/r$ and $r_H$ is the parametrized location of the event horizon. Thus, there are four metric components we are solving for, namely the set $\{F_0, F_1, F_2, \omega\}$, which are all functions of both $r$ and $\theta$, and the spacetime possesses the isometries we impose, admiting two Killing fields, $\xi=\partial_t$ and $\eta=\partial_\phi$.

\subsection{Ansatz for the scalar fields}

Fundamentally, in these theories, the scalar fields describe spin-0 gravitons. Nevertheless, when cast in the Einstein frame they take a very similar form of a bosonic matter field minimally coupled to gravity in general relativity. As discussed in our previous work \cite{Collodel:2019uns}, such solitons must depend explicitly on time for stability reasons, and on the axial coordinate if they are meant to be rotating. In the presence of a null hypersurface, these conditions become even more stringent due to the no hair theorems which we can circumvent if the matter fields (Einstein frame) do not share the spacetime isometries. In the $N=2$ theories we are interested in, the scalar field must take the following form
\begin{equation}
\label{ans}
\left\{\varphi^1,\varphi^2\right\}=\left\{\psi(r,\theta)\cos(\omega_st+m\phi),\psi(r,\theta)\sin(\omega_st+m\phi)\right\},
\end{equation}
hence the action and the effective energy momentum tensor remain independent of time and the azymuthal coordinate, and consequently the spacetime stationarity and axisymmetry are not jeopardized. There is therefore only one fundamental scalar field $\psi$, for the system is endowed with a $U(1)$ symmetry that allows us to rotate $\varphi^1$ into $\varphi^2$ and vice versa. Moreover, because $\varphi^a(\phi=0)=\varphi^a(\phi=2\pi)$, $m$ must be an integer.

\subsection{Boundary Conditions}
The radial coordinate is transformed in order to obtain simpler conditions at the horizon and to compactify the domain and bring infinity to a finite numerical value. As in \cite{PhysRevLett.112.221101}, we make $\bar{r}=\sqrt{r^2-r_H^2}$ and adopt $x=\bar{r}/(\bar{r}+1)$ as our coordinate, and the system is solved in the radial domain $x\in\left[0,1\right]$, which gives the range $r\in\left[r_H,\infty\right)$. At the horizon, except for $\omega$, all other five functions obey the trivial Neumann boundary condition, namely 
\begin{align}
\partial_xF_0\rvert_{x=0}=0,\qquad \partial_xF_1\rvert_{x=0}=0,\qquad \partial_xF_2\rvert_{x=0}=0,\qquad \partial_x\psi\rvert_{x=0}=0.
\end{align}
Regularity requires that
\begin{align}
\omega\rvert_{x=0}=-\frac{r_H\omega_s}{m},
\end{align}
which also translates to the critical value of the natural frequency $\omega_s$ that allows for bound solutions of scalar clouds on a Kerr background, as all other frequencies would be complex rendering the system unstable, i.e. the scalar cloud would decay as it falls into the black hole ($\operatorname{Im}\omega_s>0$), or superradiant instability arises ($\operatorname{Im}\omega_s<0$). Note that, from our chosen form of the metric (\ref{metric}), the horizon angular velocity is given by
\begin{align}
\label{homega}
\Omega_H&=\frac{\omega(x=0)}{r_H} \nonumber \\
        &=-\frac{\omega_s}{m},
\end{align}
for which reason the nontriviality of these scalar fields makes them \emph{synchronized hair} \cite{PhysRevLett.112.221101}. Furthermore, there is a Killing field $\chi^\mu=\xi^\mu+\Omega_H\eta^\mu$ which is null at the horizon and generates the flow along which the scalar field is dragged, for $\mathcal{L}_\chi\varphi^1=\mathcal{L}_\chi\varphi^2=0$, and there is no flux into the black hole.

The spacetime is asymptotically flat, and the field $\psi$ vanishes at infinity,
\begin{align}
F_0\rvert_{x=1}=0,\qquad F_1\rvert_{x=1}=0,\qquad F_2\rvert_{x=1}=0,\qquad \omega\rvert_{x=1}=0, \qquad \psi\rvert_{x=1}=0.
\end{align}
The asymptotic behavior of $\psi$ (because of asymptotic flatness) is precisely the same as it is for solitons in the absence of a horizon,
\begin{align}
\psi(r\sim\infty)\propto\frac{1}{r}\exp{\left(-\sqrt{\mu^2-\omega_s^2}r\right)},
\end{align}
so that that the constraint on the frequency, $\omega_s^2\leq\mu^2$, still holds and the scalar field becomes trivial at the equality.

On the symmetry axis, regularity and axisymmetry require
\begin{align}
\partial_\theta F_0\rvert_{\theta=0}=0,\qquad \partial_\theta F_1\rvert_{\theta=0}=0,\qquad \partial_\theta F_2\rvert_{\theta=0}=0,\qquad \partial_\theta\omega\rvert_{\theta=0}=0, \qquad \psi\rvert_{\theta=0}=0,
\end{align}
while on the equatorial plane, for even parity systems, we have
\begin{align}
&\partial_\theta F_0\rvert_{\theta=\pi/2}=0,\qquad \partial_\theta F_1\rvert_{\theta=\pi/2}=0,\qquad \partial_\theta F_2\rvert_{\theta=\pi/2}=0,\qquad \partial_\theta\omega\rvert_{\theta=\pi/2}=0,\qquad \partial_\theta\psi\rvert_{\theta=\pi/2}=0.
\end{align}

\subsection{GLOBAL CHARGES}

The ADM mass and angular momentum can be extracted asymptotically from the metric functions' leading terms at infinity,
\begin{equation}
\label{asympmj}
M=\frac{1}{2}\lim_{r\rightarrow\infty}r^2\partial_rF_0, \qquad J=\frac{1}{2}\lim_{r\rightarrow\infty}r^2\omega,
\end{equation}
and can also be calculated through the Komar integrals defined with the spacetime Killing vector fields, which after explicitly breaking down to the contributions given by the hole itself ($M_H, J_H$) and the scalar hair ($M_\psi, J_\psi$) give
\begin{equation}
\label{komar}
M=\underbrace{2\int_\mathcal{H} n^\mu\sigma^\nu\nabla_\mu\xi_\nu dA_H}_{M_H} + \underbrace{2\int_{\Sigma\backslash\mathcal{H}} n^\mu\nabla_\nu\nabla_\mu\xi^\nu dV}_{M_\psi},\qquad J=\underbrace{-\int_\mathcal{H} n^\mu\sigma^\nu\nabla_\mu\eta_\nu dA_H}_{J_H} \underbrace{-\int_{\Sigma\backslash\mathcal{H}} n^\mu\nabla_\nu\nabla_\mu\eta^\nu dV}_{J_\psi},
\end{equation} 
where $\mathcal{H}$ is the horizon surface, $dA_H$ its area element and $\sigma^\mu$ a spacelike vector perpendicular to it such that $\sigma^\mu\sigma_\mu=1$. Similarly, $\Sigma\backslash\mathcal{H}$ is a spacelike asymptotically flat hypersurface bounded by the horizon, $dV$ is its volume element and $n^\mu$ is a timelike vector normal to it such that $n^\mu n_\mu=-1$. The metric (\ref{metric}) implies that $\sigma^\mu=(0,\sqrt{\mathcal{N}}e^{-F_1},0,0)$, $n^{\mu}=(\xi^\mu+\omega/r\eta^\mu)/\left(\sqrt{\mathcal{N}}e^{F_0}\right)$, $dA_H=e^{F_1+F_2}r^2\sin\theta d\theta d\phi$ and $dV=\sqrt{-g/\mathcal{N}}e^{-F_0}drd\theta d\phi$. Unveiling the scalar field's contributions in explicit form, yields 
\begin{equation}
\label{intmj}
M_\psi=\frac{1}{8\pi}\int_{r_H}^\infty\int_0^\pi\int_0^{2\pi}\left(2\frac{\omega_s}{m}\mathfrak{j}-U\right)\sqrt{-g}drd\theta d\phi, \qquad J_\psi=\frac{1}{8\pi}\int_{r_H}^\infty\int_0^\pi\int_0^{2\pi}\mathfrak{j}\sqrt{-g}drd\theta d\phi,
\end{equation}
where we define the angular momentum density
\begin{equation}
\mathfrak{j}=2m\frac{\left(m\omega+\omega_sr\right)\Omega^2\psi^2}{Ne^{2F_0}r}.
\end{equation}

The conserved current \eqref{nc} has an associated conserved Noether charge which is obtained by projecting it along the future directed vector field $n^\mu$ and integrating over the whole $\Sigma\backslash\mathcal{H}$ domain, 
\begin{equation}
\label{nnc}
Q = \int_{\Sigma\backslash\mathcal{H}}  \tilde{j}_\mu n^\mu dV, 
\end{equation}
and by applying the same substitutions we did for the Komar integrals, we arrive at
\begin{equation}
\label{nnc2}
Q = \frac{1}{8\pi}\int_{r_H}^\infty\int_0^\pi\int_0^{2\pi}\frac{\mathfrak{j}}{m}\sqrt{-g}drd\theta d\phi,
\end{equation}
and we observe that even in the presence of a horizon, the angular momentum due to the scalar field remains quantized as it happens for solitons, $J_\psi=mQ$. Following the approach done in \cite{Herdeiro:2014goa}, we define a normalized charge $q=J_\psi/J\in [0,1]$ which serves as a measure of how hairy the black hole is.

The hole's temperature and entropy are given by $T_H=2e^{F_0-F_1}/r_H$ and $S=A_H/4$, respectively. Together with the above relations, after manipulating the expression for $M_H$ we arrive at the Smarr Law for these black holes,
\begin{equation}
\label{smarr}
M=2T_HS-2m\Omega_HJ_\psi+M_\psi.
\end{equation}

These three charges $(M, J, Q)$ are frame independent, and therefore the same for all conformal factors $A(\psi)$. The horizon area $A_H$ is not, however, and is written in the Einstein frame, but the entropy remains invariant due to the hidden gravitational constant.

\section{Results} \label{sec:Results}

\subsection{Numerical Setup}\label{sec:NumericalSetup}

The set of five coupled nonlinear PDE are solved with aid of the program package FIDISOL \cite{SCHONAUER1990279}, which employs a finite difference discretization allied with a relaxed Newton scheme and different linear solvers. We investigate five theories, each described by a different value of the Gauss curvature, $\kappa\in\{-5,-1,0,1,5\}$, as we did for solitons in \cite{Collodel:2019uns}. The solutions scale with the scalar field mass $\mu$, and all other constants are input parameters, namely $\{\kappa, \omega_s, m, r_H \}$. Since we are interested in the most fundamental spinning black holes in these theories we restrain our analysis to the $m=1$, as higher values correspond to more excited states which are quantum mechanically more unstable, but stress that there is in principle no theoretical upper bound for the winding number. In what concerns the parameter space of these solutions, higher $m$ would allow for solutions with greater mass and angular momentum, see e.g. \cite{Delgado:2019prc}. Wherever the conformal factor is needed we choose $\beta=-6$ for illustrative purposes, as we must add that for these calculations, any plausible value for this parameter produces very similar outcomes. We start with a solitonic solution ($r_H=0$) near its limiting value $\omega_s\sim\mu$ and \emph{add a small black hole at the center} by slightly increasing the horizon radius and then modifying the boundary conditions accordingly. For any fixed $r_H$, we vary $\omega_s$ to obtain a new solution belonging to this parameter family curve until we reach the boundary of the existence domain, or until convergence within an acceptable error range is no longer possible. The accuracy of each solution is tracked through the maximum component error calculated by FIDISOL, and by comparing the global quantities obtained asymptotically, eq. (\ref{asympmj}), with those from direct integration, eq. (\ref{komar}), and verifying the Smarr relation, eq. (\ref{smarr}). In all cases we kept the relative errors under $10^{-4}$. For a detailed and thorough description on how similar solutions are built, we refer the reader to \cite{Herdeiro_2015}.

\subsection{Domain of Existence}

\begin{figure}[htp]
\centering
\includegraphics[width=.45\textwidth]{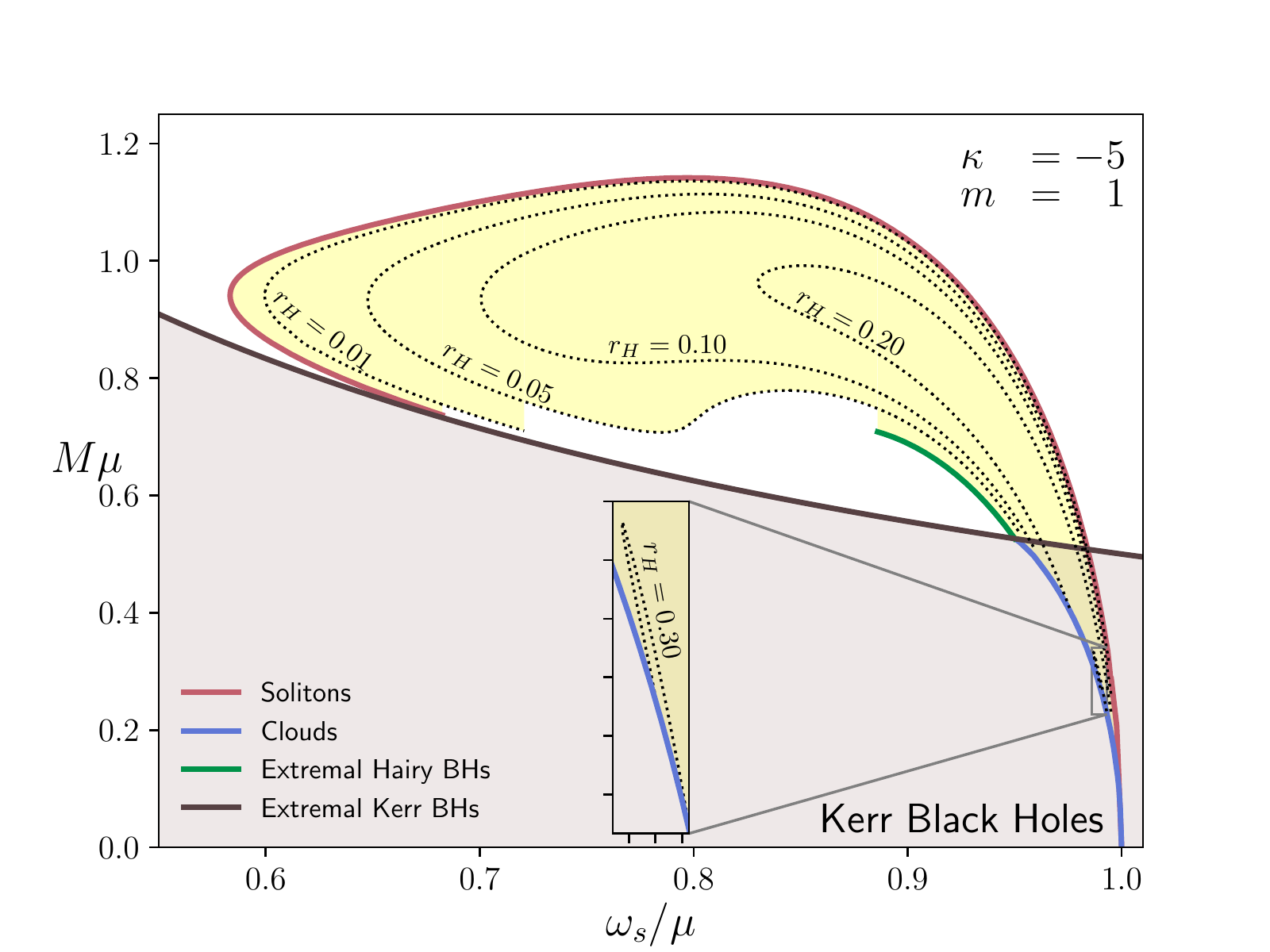}\quad
\includegraphics[width=.45\textwidth]{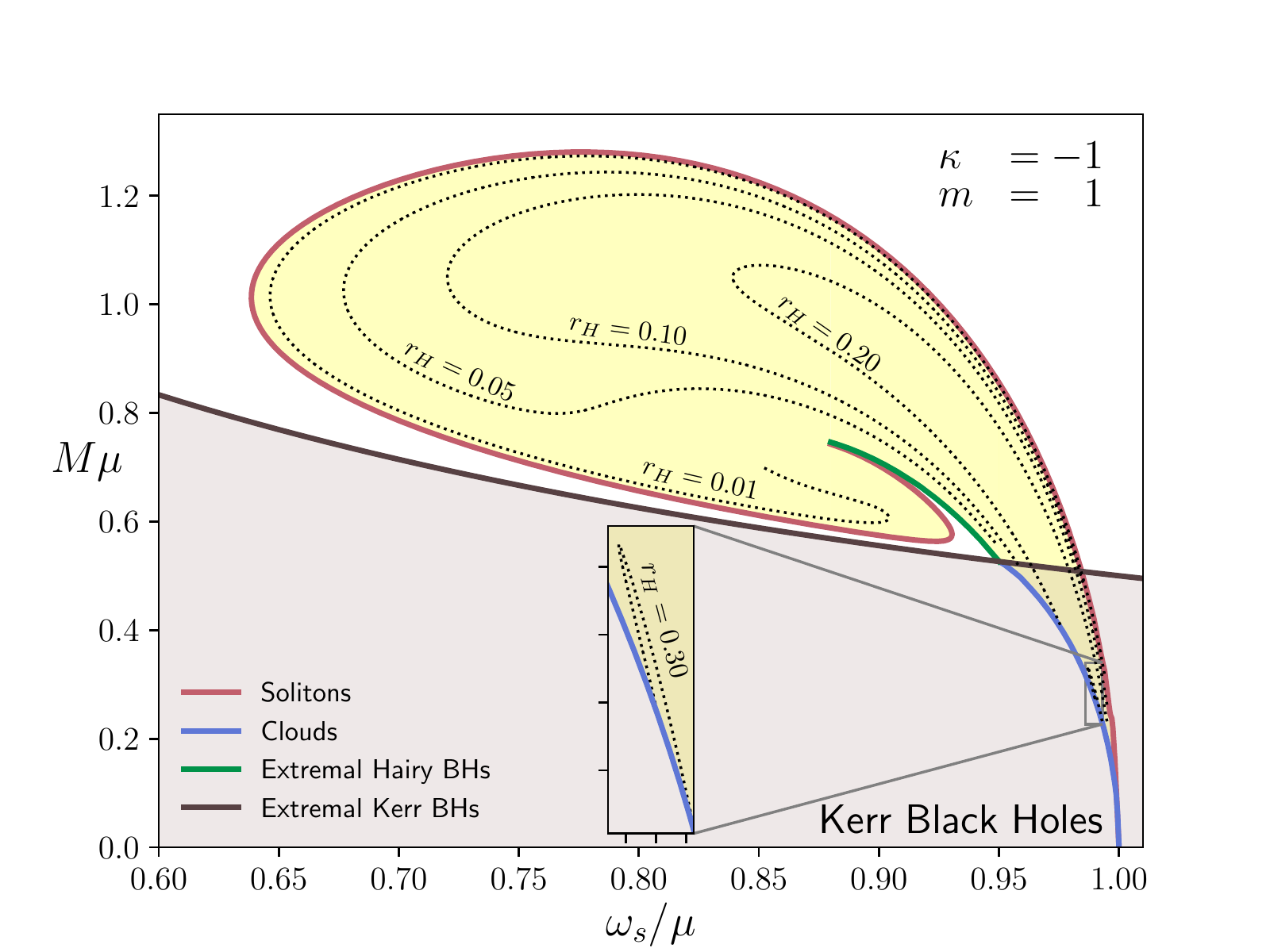}\quad

\medskip

\includegraphics[width=.45\textwidth]{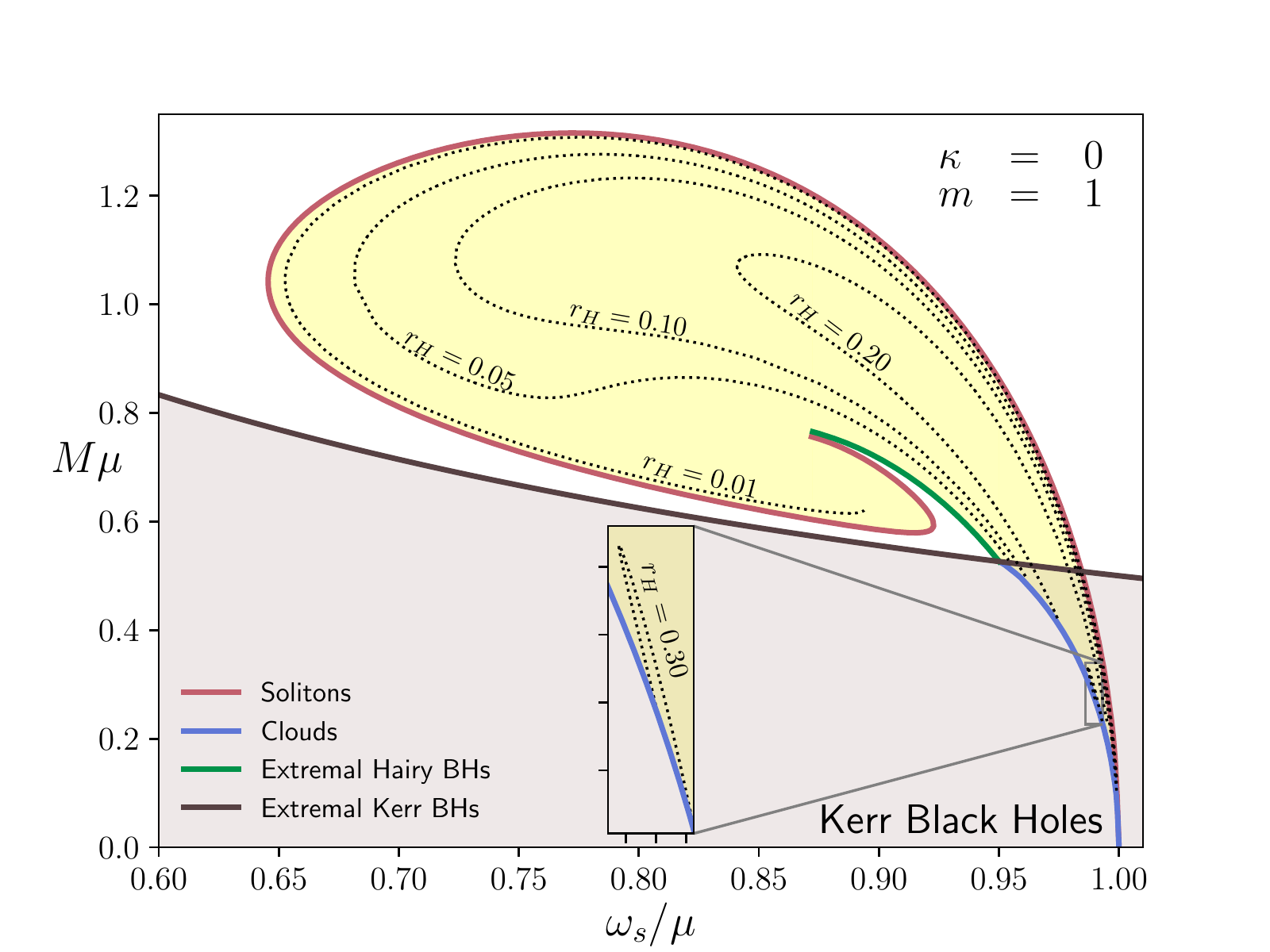}

\medskip

\includegraphics[width=.45\textwidth]{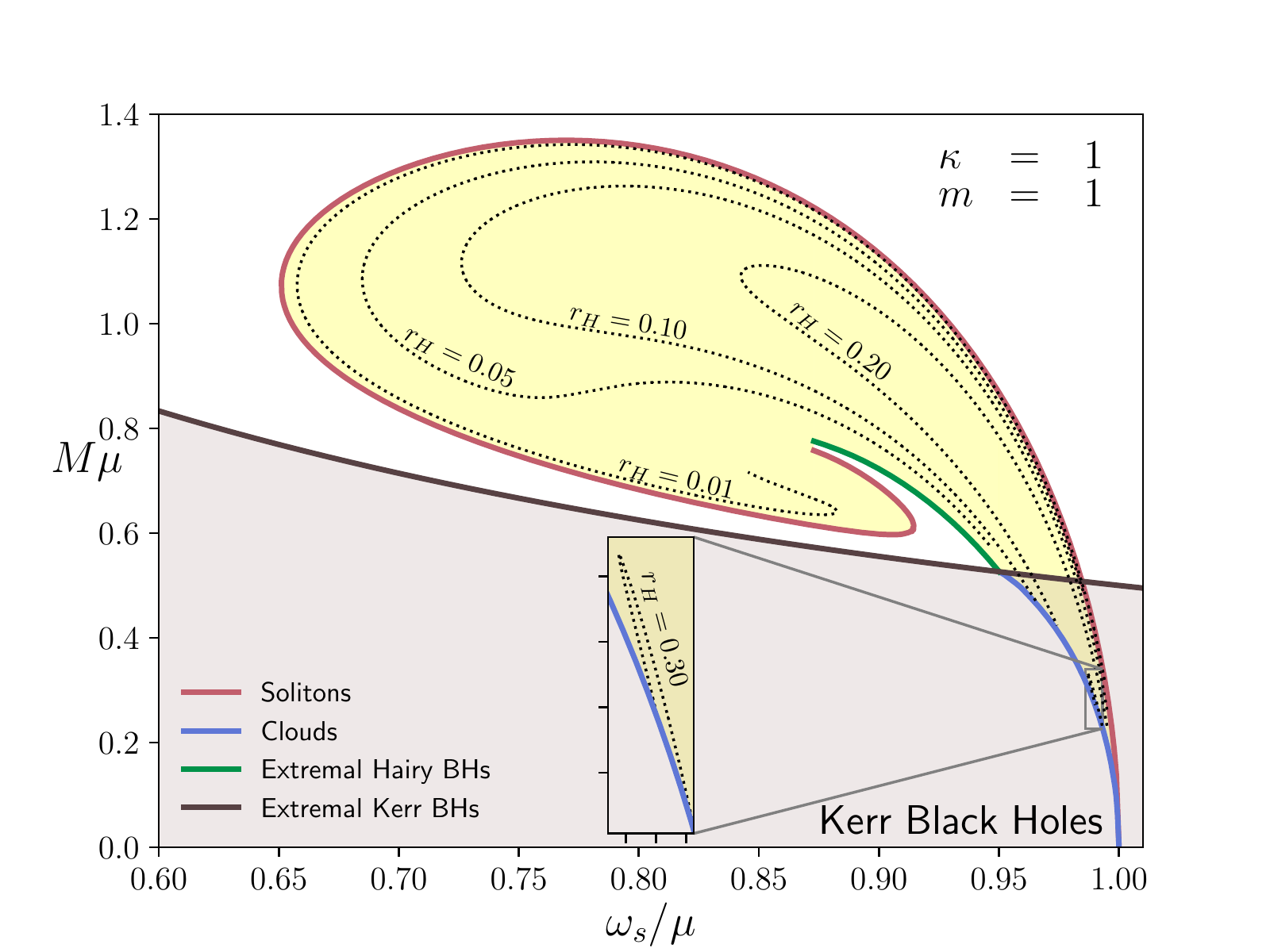}\quad
\includegraphics[width=.45\textwidth]{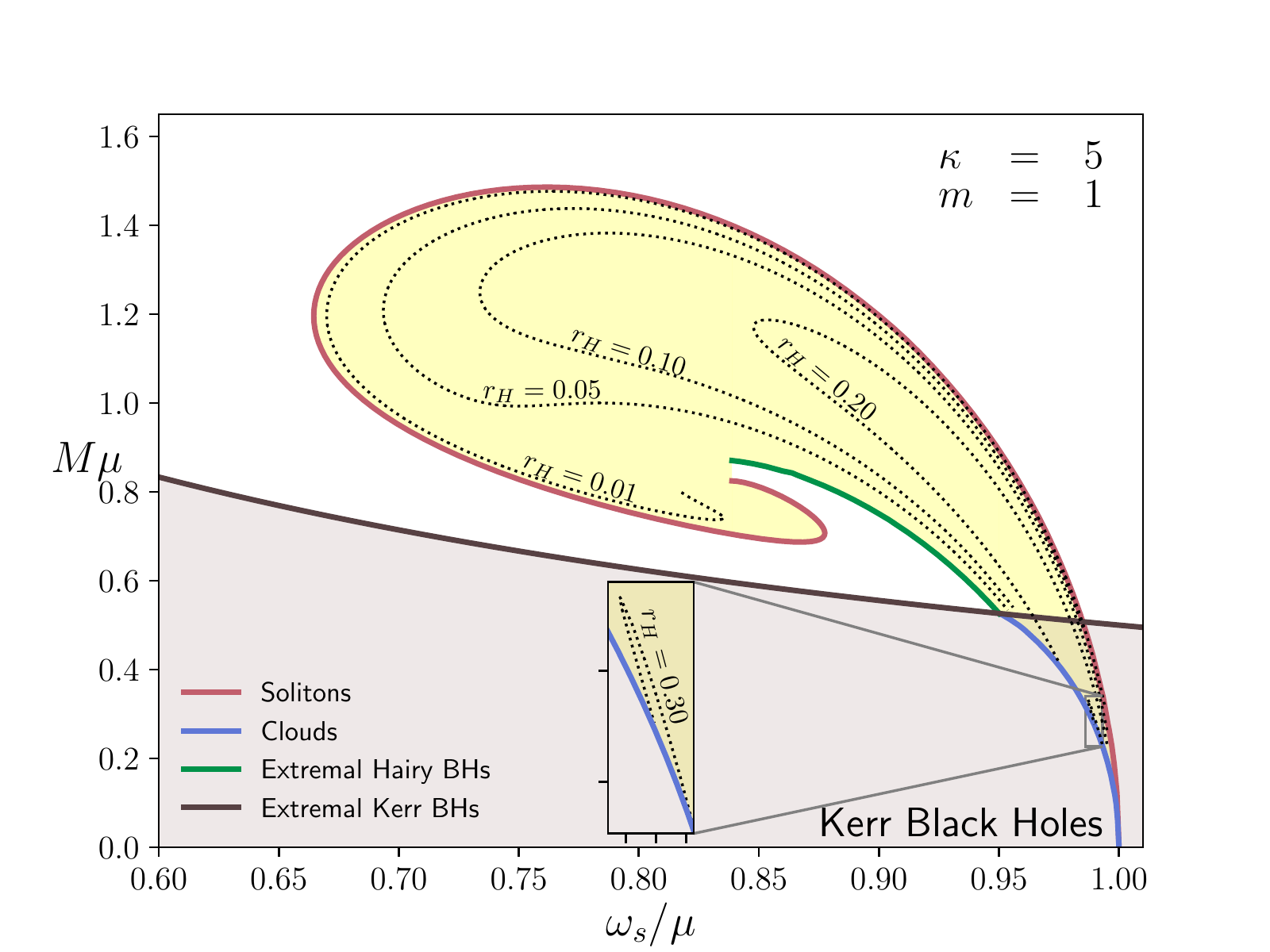}

\caption{Mass vs. $\omega_s/\mu$ diagram for five different Gauss curvatures. Hairy black hole solutions exist within the yellow area enclosed by three boundaries of distinct objects: solitons (red), extremal black holes (green) and clouds (blue), where the latter are bound solutions of the KGE equation on a Kerr background. The gray shaded region is where Kerr black holes are found and is bounded by extremal Kerr solutions (black) where the spin parameter gets saturated, $j=J/M^2=1$. In all cases the winding number of the scalar field is taken to be one $m=1$, which gives the most fundamental state for rotating solutions.}
\label{fig:mxo}
\end{figure}

There are three distinct boundaries that enclose the domain of existence of hairy black holes in these theories, for each fixed $m$ and $\kappa$, which are qualitatively the same as for the Herdeiro-Radu black holes (HRBH), and each of these boundaries have either a fixed normalized charge $q$, or a fixed horizon radius $r_H$, or both. They are:
\begin{itemize}
\item Solitons ($q=1$, $r_H=0$).
\item Clouds ($q=0$).
\item Extremal Hairy Black Holes ($r_H=0$).
\end{itemize}
Furthermore, because the KGE equation (\ref{kge}) for this system allows for trivial scalar field solutions in vacuum, Schwarzschild and Kerr black holes are also stationary solutions of these theories. In Fig. \ref{fig:mxo}, we display in yellow background color the domain of existence contained within the boundaries described above for the five theories we are considering. Kerr black holes exist in the shaded region under the black curve. Extending our previous work on rotating solitons  in TMST \cite{Collodel:2019uns}, we have built extra solitonic solutions (red curves) which start at $\omega_s/\mu=1$ and form a curve in this parameter space that has turning points. The first one is at the solution of minimum frequency. For positive $\kappa$ this turning point moves to larger $\omega_s/\mu$, but on the other hand the maximum mass attained is also bigger. Negative $\kappa$, though, lead to smaller value of $\omega_s/\mu$ at the first turning point (e.g. for $\kappa=-5$, $\omega_s/\mu$ drops below $0.6$ at the first turning point) and hairy black hole exist for a larger interval of values for the horizon's angular velocity. 

After the first turning point, $\omega_s/\mu$ of the solitonic red curve solutions start to increase until reaching a second turning point. After this the red curve starts winding up on itself in a place referred to as the \emph{inspirilling region} that is known to be notoriously challenging numerically. We therefore only display solutions before the third turning point. As one advances along the solitonic curve towards the spiral, the lapse function grows indefinitely, and it seems indeed that this curve meets the one for extremal hairy black holes (green curve), for which $r_H=0$, at the center of the spiral. These extremal black holes start at this point with $q=1$, and the following point have decreasing charge until the solution coincides with the line for extremal Kerr black holes (black line where $M=0.5/\omega_s$ and $q=0$). This marks the starting point of the cloud solutions (blue curve), namely marginally bound solution of eq. (\ref{kge}) on a Kerr background.

The above described behavior is observed for all of the positive as well as the slightly negative $\kappa$. With the decrease of this parameter, though, e.g. the $\kappa=-5$ case shown in the first panel in Fig. \ref{fig:mxo}, the convergence stops well before the desired region as the scalar field profile becomes quite steep and it is difficult to tell where the solitons should meet the extremal holes. In all cases there is a non-uniqueness region where for a single horizon angular velocity and mass, two solutions exist: a Kerr solution, and a hairy one.

Generally, for a fixed $r_H$ sequence of hairy black holes (depicted with doted lines), one end-point of the sequences  sits at the trivial Minkowski spacetime ($\omega_s/\mu=1$), while the other is at a point on the cloud curve. This behavior changes for very large radii, e.g. $r_H=0.3$ (zoomed in inset), where both extremes meet at the cloud.  The higher the value of the horizon radius is, the more compact is its support on $\omega_s$, which is to be expected, as the smaller the radius gets, the closer the curve falls to the solitonic one. 

In the five cases of study, the cloud curves (blue solid line) are the same, and so are the scalar cloud solutions within the adopted precision. The theoretical justification for this observations if the following. On a Kerr background, the solutions of the KGE equation are given by an infinitesimal field, and the term involving $\kappa$ which differentiates the theories is basically a small perturbation on it. To find meaningful deviations, one needs to increase the absolute value of $\kappa$ by two orders of magnitude. The extremal solutions, though, differ non-negligibly for various $\kappa$ even though they all have the same ending point on the extremal Kerr  sequence of solutions. 

\begin{figure}[htp]
\centering
\includegraphics[width=.6\textwidth]{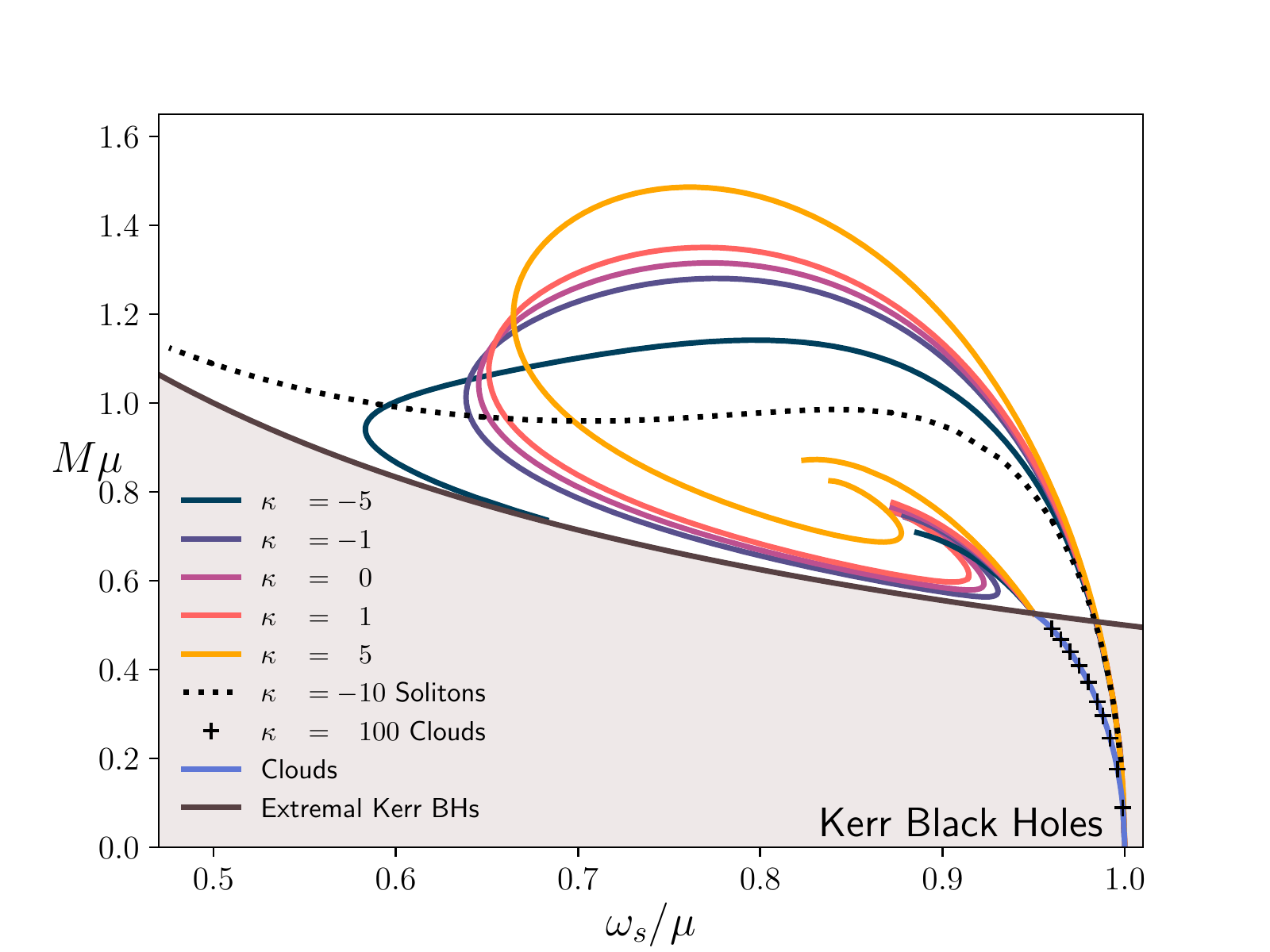}
\caption{Mass vs. $\omega_s/\mu$ diagram for the boundaries (the joint curves of solitons, extremal black holes and clouds) of the five different models studied, $\kappa={-5, -1, 0, 1, 5}$, overlapped. The cloud curves all fall exactly onto one another and are therefore colored blue. In addition, the plus signs are cloud solutions for $\kappa=100$, which is a case not studied here, but exemplify that much higher values of the Gauss curvature are required to see deviations from this line. The dotted curve is a set of solitonic solutions ($q=1$) for $\kappa=-10$, where no turning point was achieved.}
\label{fig:mxoall}
\end{figure}

In order to better appreciate the distinction in the existence diagram between these different models, we display in Fig. \ref{fig:mxoall} the domain boundaries for all five cases. For purely illustration purposes, we add some solution points of clouds in the $\kappa=100$ case, which still overlap the cloud boundary of the five models here investigated. The challenge in continuing the curves for $\kappa=-5$ could have a more fundamental reason if there is an intrinsic change of behavior of the solutions sets as one decreases further the Gauss curvature. Therefore, we add a set of solitonic solutions curve for a theory with $\kappa=-10$, where we observe that by following the curve from the trivial solution ($\omega_s/\mu\sim 1$), it does not reach a global maximum followed by a turning point. Instead, it features local maximum, decreases to a local minimum and starts growing again with decreasing natural frequency. The curve stops where convergence cannot be achieved anymore, and this happens before any turning point appears.

\subsection{Properties}

\begin{figure}[htp]
\centering
\includegraphics[width=.4\textwidth]{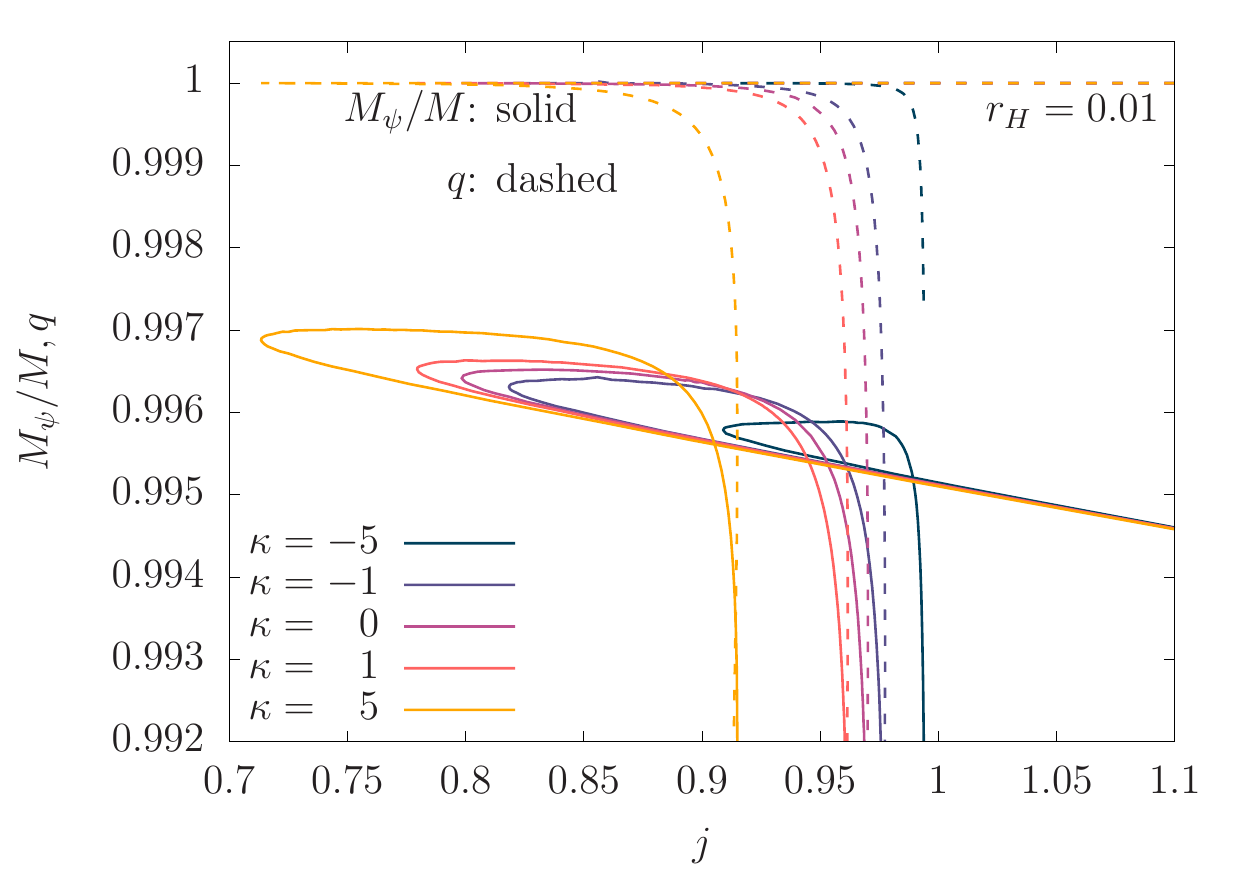}
\includegraphics[width=.4\textwidth]{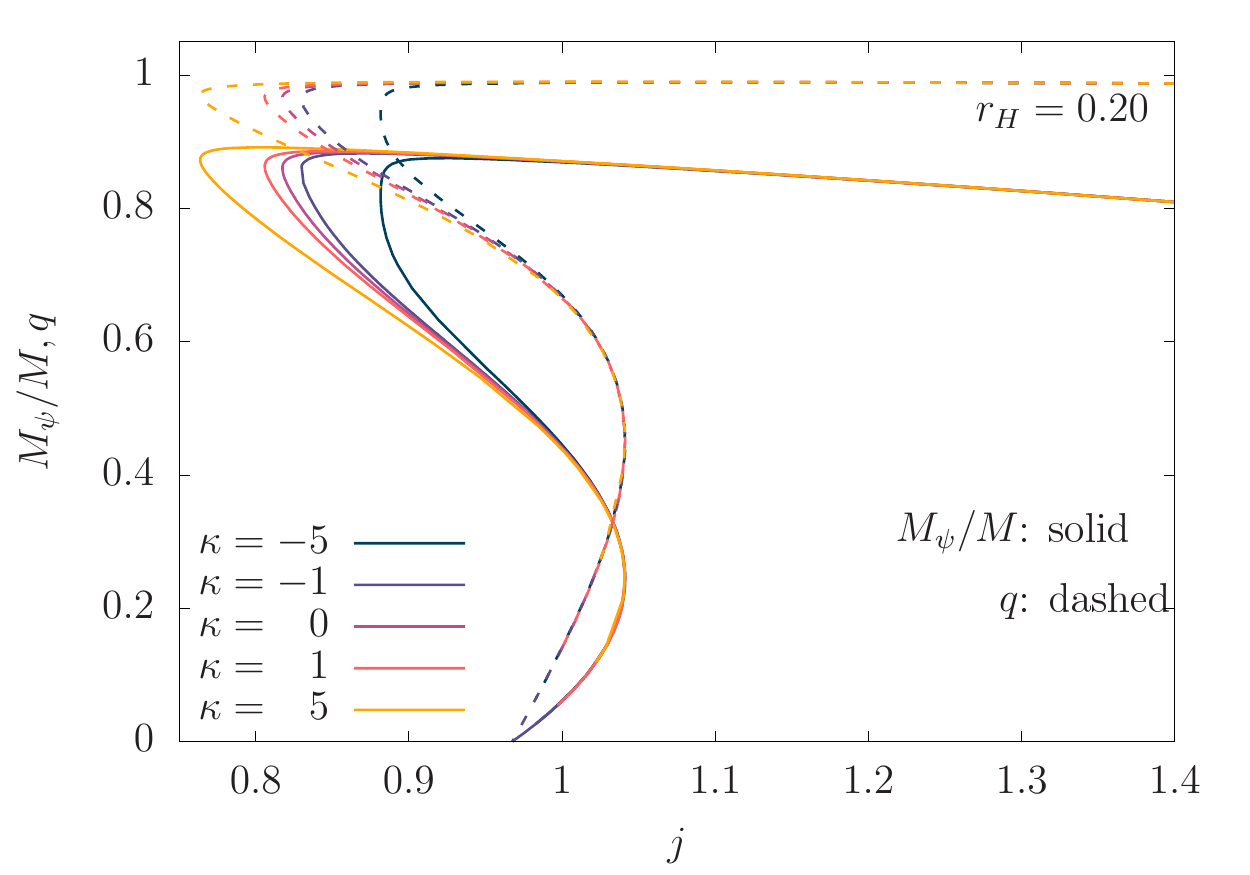}
\caption{Mass ratio (solid) and normalized charge (dashed) against the spin parameter for five different values of $\kappa$. With solid lines the $M_\psi/M(j)$ dependencies are depicted while the $q(j)$ dependencies are plotted with dashed lines. \emph{Left:} Small horizon radius, $r_H=0.01$. \emph{Right:} Large horizon radius, $r_H=0.2$. Different interval domains on the vertical axis are chosen as for very small radii the solutions depart steeply from the clouds (no meaningful change in $j$ as the hair grows more massive) and most of the points lie in a region where the ratio is very close to one, as such objects are much closer to solitons than to Kerr black holes. }
\label{fig:momxj}
\end{figure}

In order to quantify how hairy a solution is, we may either turn to the normalized charge $q$ which also tells how much angular momentum is stored in the hair, or to the ratio $M_\psi/M$ which shows how much of the total mass is in fact contributed by the scalar hair. By such measures, the pure solitonic solutions are of course the most hairy, since they possess no holes. In Fig. \ref{fig:momxj} we present $q$ and $M_\psi/M$ as functions of $j=J/M^2$ for all five values of $\kappa$ and for both a big and a small horizon radius. Beyond the Kerr limit ($j=1$), all curves for different $\kappa$ fall very close to each other and become almost indistinguishable and, as expected those solutions are very hairy. There is, however a big difference in the behavior of the two quantities analyzed, i.e. the solutions in this region are not the ones with the greatest mass ratio $M_\psi/M(j)$ of all, but the ones with the greatest normalized charge. This feature can be understood by pondering on the extreme limits of these curves. As we move to the right in the figure and extrapolate the displayed region, we approach the critical value of $\omega_s=\mu$, which corresponds to the trivial solution where $M=0$, spacetime is flat and this ratio measure $M_\psi/M(j)$ loses meaning. However, because up until this point the centered hole still exists, there is always some mass attributed to it and the mass ratio falls, even if ever so slowly, while the angular momentum ratio appears to be constant. On the other end of the curves both $M_\psi/M(j)$ and $q$ drop sharply until we reach the clouds on a Kerr background where both quantities are zero. In this limit the lines for different $\kappa$ as also indistinguishable.

In comparing the different theories, one can notice that significant deviations between different $\kappa$ are observed only close to the minimum of $j$.  This is well justified, since the two ending points of the sequence of constant $r_H$ solutions, as described in the previous section, are practically independent of $\kappa$ for the considered range of Gauss curvatures. In addition, the larger the Gauss curvature is, the smaller is the minimum value of $j$ achieved for a certain fixed horizon radius. For the smallest curvature investigated, $\kappa=-5$, all objects present $j>0.8$, which justifies the difficulty in finding solutions in a region of the parameter space where they approach the Kerr black hole. Following this line of thought, one should ask if for smaller values of $\kappa$ - and which - clouds cease to exist, meaning all solutions would feature $j>1$ and the KGE on a Kerr background would in turn yield no bound solutions. Since this region is notoriously difficult to calculate numerically, we leave these studies to a future publication, but indeed it is expected that such limit would exist. We also note that the maximum mass ratio for a given radius increases with the Gauss curvature, and that for very small radii, the solution depart quite steeply from Kerr to values of $M_\psi/M$ very close to one, which would correspond to the solitonic case.

\begin{figure}[htp]
\centering
\includegraphics[width=.4\textwidth]{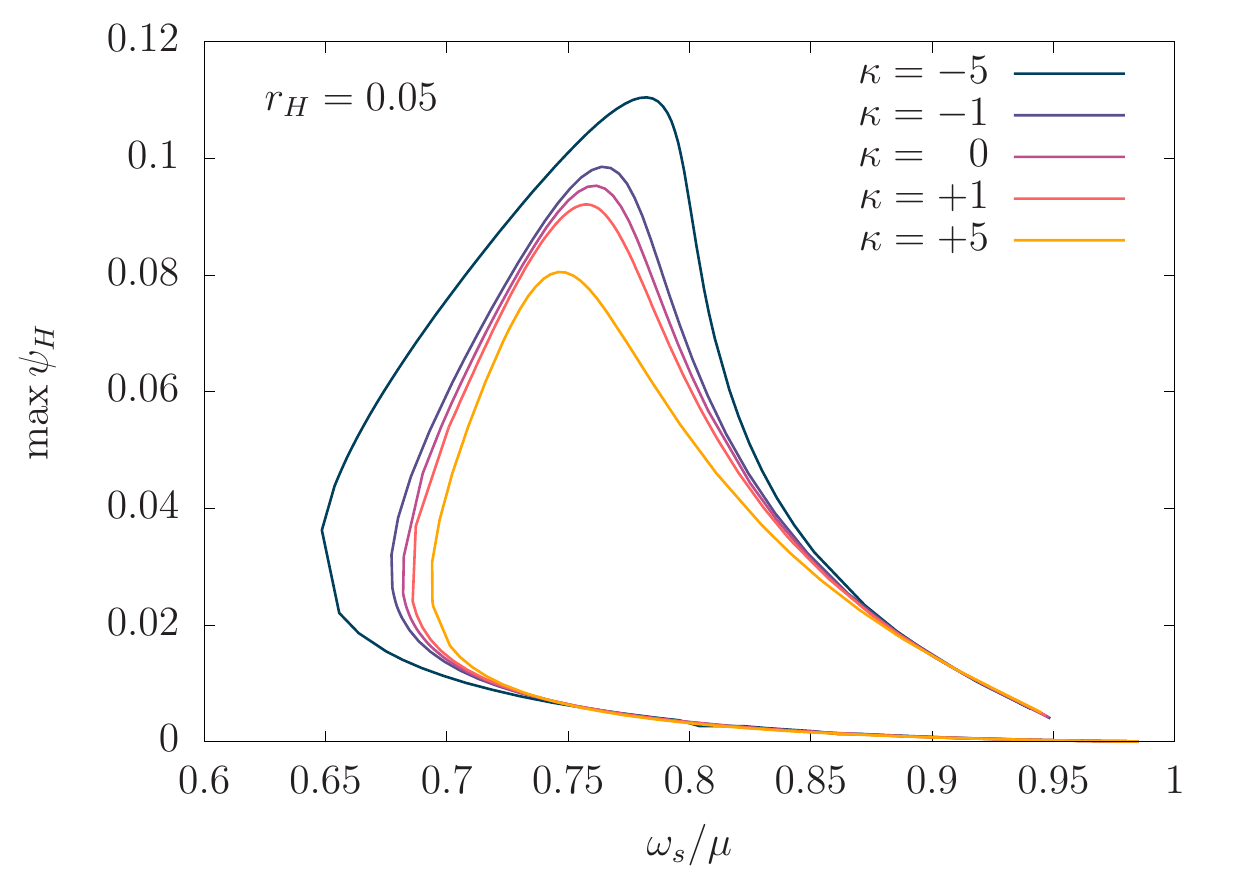}
\includegraphics[width=.4\textwidth]{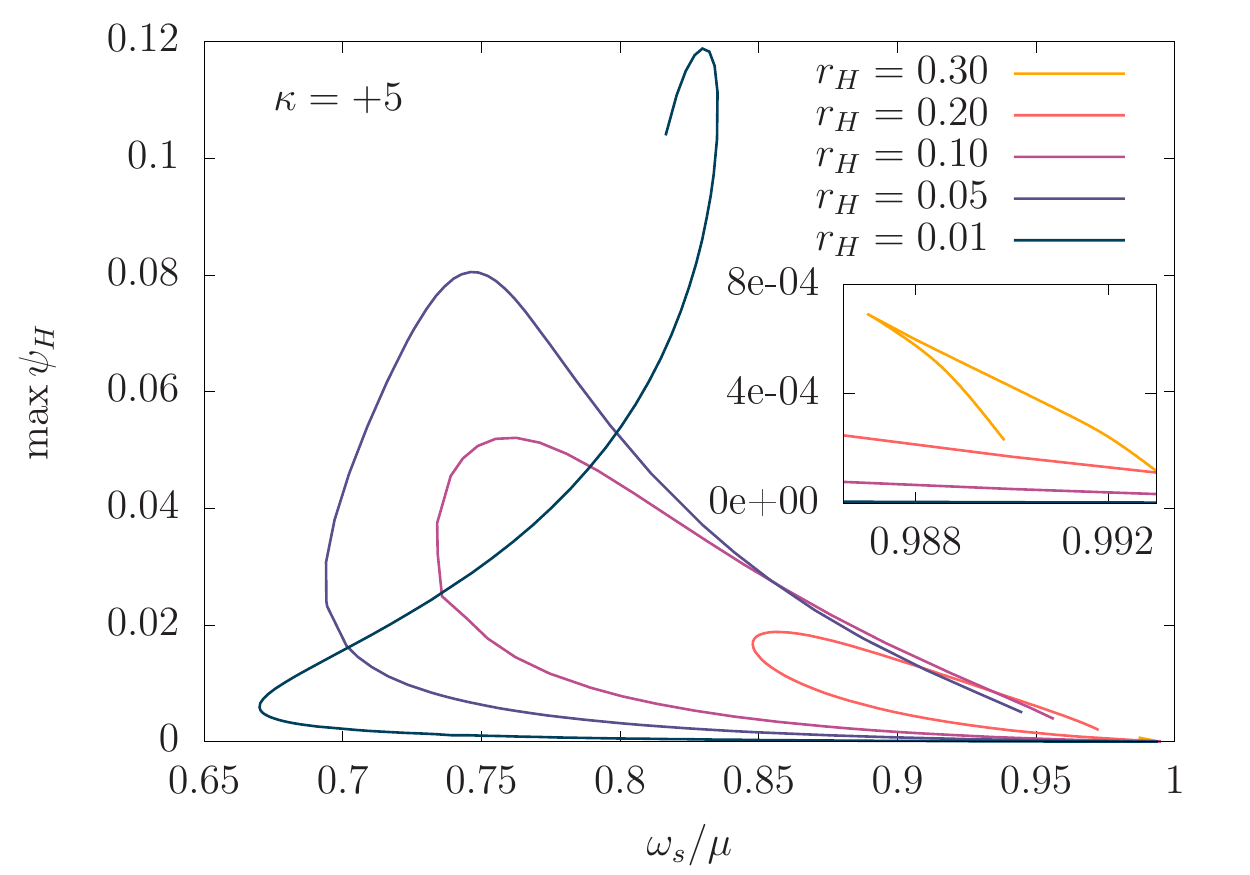}
\caption{Maximum value taken by the scalar field at the $\psi$ at the horizon as a function of $\omega_s/\mu$. \emph{Left:} Fixed horizon radius at $r_H=0.05$ and five different values of $\kappa$. \emph{Right:} Fixed $\kappa=5$ and five different values of $r_H$. As either the Gauss curvature or the horizon radius increases in value, the maximum value of $\psi$ at the horizon decreases.}
\label{fig:mp}
\end{figure}

The maximum value the scalar field assumed at the horizon is depicted in Fig. \ref{fig:mp} for the five values of $\kappa$ and $r_H=0.05$ (left), and for five values of $r_H$ and $\kappa=5$ (right), for illustration purposes. According to the boundary conditions, the field disappears on the poles and is generally quite small along this surface. Moreover, this quantity decreases in value as either the Gauss curvature or the horizon radius increases. The latter is intuitive, as for a certain (unknown) value of $r_H$ hairy solutions cease to exist and one is left with bald Kerr black holes, but we recall as seen above that as $\kappa$ increases, so does the maximum value of the ratio $M_\psi/M$ and one could naively expect the scalar field to take larger values everywhere, including the horizon.

\begin{figure}[htp]
\centering
\includegraphics[width=.4\textwidth]{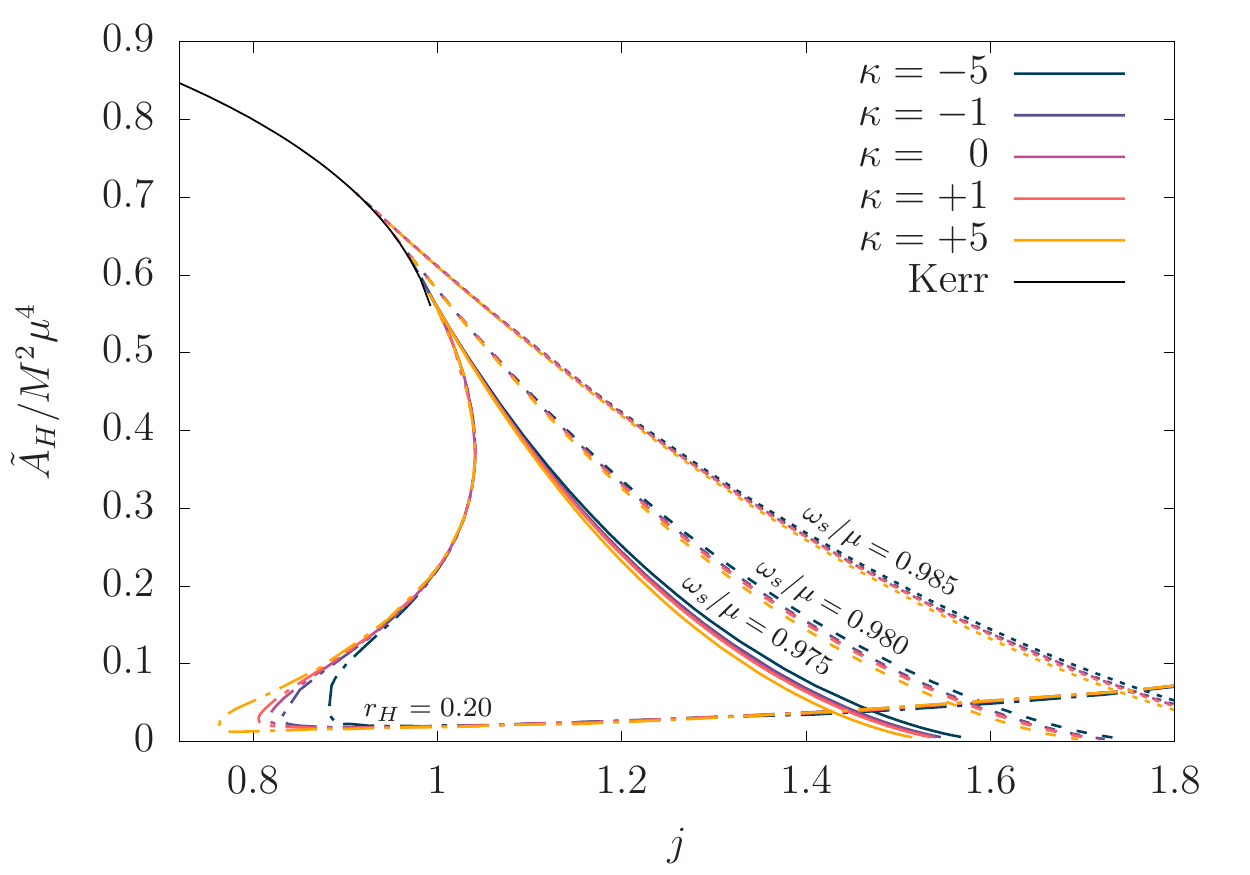}
\includegraphics[width=.4\textwidth]{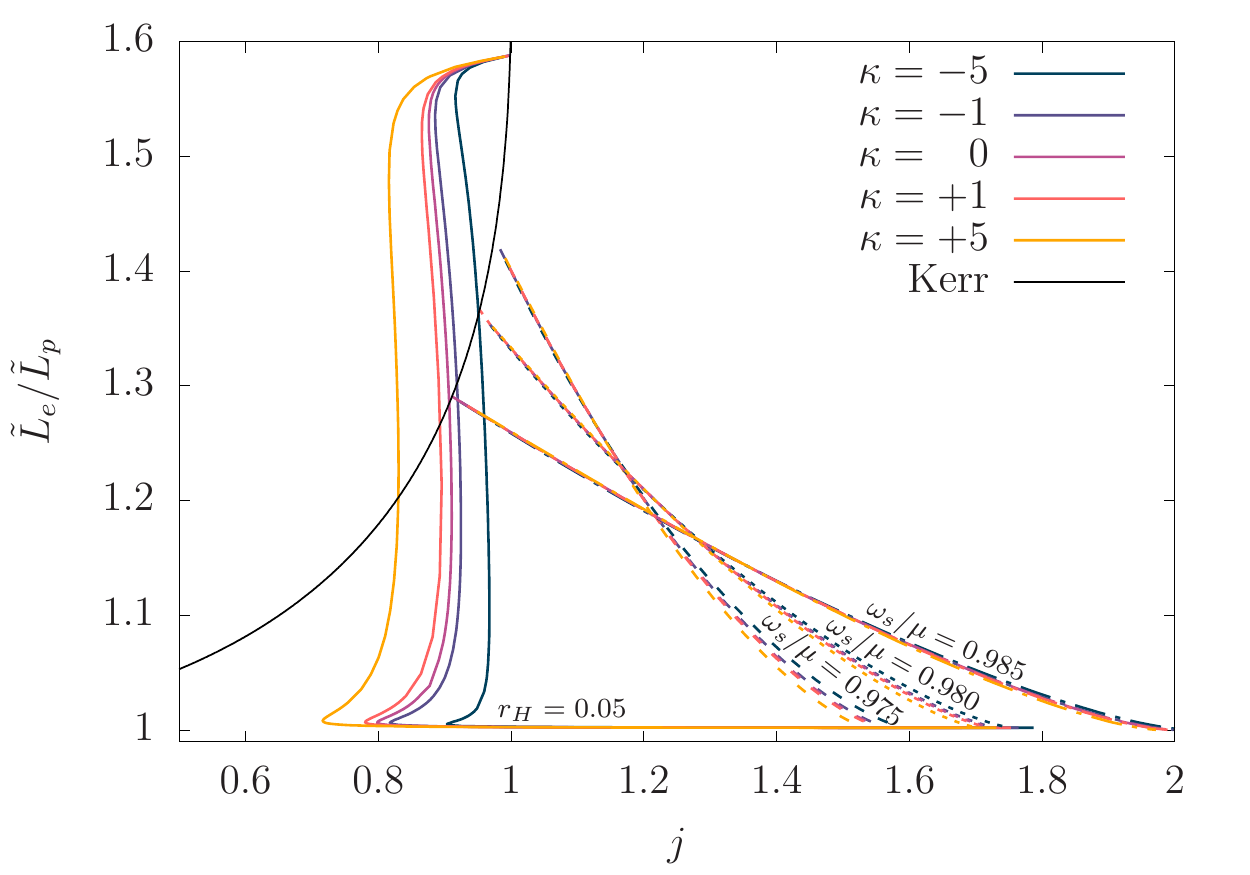}
\caption{\emph{Left:} Horizon reduced area over the mass squared against the spin parameter for the five different theories considered, exemplified with fixed $r_H=0.2$ and three other sets of fixed horizon angular velocity. \emph{Right:} Horizon deformation factor against the spin parameter for the same sets of $\omega_s/\mu$ and for $r_H=0.05$.}
\label{fig:ahxj}
\end{figure}

The reduced horizon area is shown on the left panel of Fig. \ref{fig:ahxj} for $\beta=-6$, although because the scalar field is small there and the form of the conformal factor we adopted in eq. (\ref{conffactor}), no apparent difference can be seen from sensibly changing $\beta$. Here it is also seen that, as the five theories are basically indistinguishable at the clouds, the difference in the reduced horizon area only becomes noticeable for more hairy solutions, and generally for fixed $j$ and $\omega_s/\mu$, the reduced area decreases with $\kappa$. In the non-uniqueness region $j\leq 1$, the hairy solutions approach the Kerr line from above, displaying a greater horizon area for the same mass and angular momentum. There are, however, other solutions along lines of constant radius, further away in the parameter space (around the solutions of maximum mass within the curve)  which also lie in this domain but feature smaller horizon areas than Kerr BHs for a same pair of mass and angular momentum. One can approach this fact intuitively by thinking of a very massive solitonic solution ($r_H=0$) in a region where $j<1$ - and therefore within the Kerr bound - to which a tiny hole can be added without effectively contributing to the ADM mass or angular momentum. Such horizon immersed in scalar matter must clearly have an area much smaller than a Kerr hole with the same global quantities.

On the right side of Fig. \ref{fig:ahxj} we display the deformation parameter, measuring how oblate the horizon is, as a function of $j$. This is given by the ratio between the value of the horizon circumference on the equatorial plane, and the one along a big circle of constant $\phi$ going through the poles,
\begin{equation}
\label{def}
\tilde{L}_e=2\pi r_H e^{F_2(r_H,\pi/2)}A\left(\psi(r_H,\pi/2)\right), \qquad \tilde{L}_p=2r_H\int_0^\pi e^{F_1(r_H,\theta)}A(\psi(r_H,\theta))d\theta,
\end{equation}
where again we used $\beta=-6$. The black line depicts the deformation of Kerr BHs, and we recall that 
\begin{equation}
\label{defkerr}
L_e^{Kerr}=4\pi M, \qquad L_p^{Kerr}=2M\int_0^\pi\sqrt{2\left(1+\sqrt{1-j^2}\right)-j^2\sin^2\theta}d\theta,
\end{equation}
and the maximum deformation occurs for the extremal Kerr, where $L_e^{Kerr}/L_p^{Kerr}=1.65$. As seen from Fig. \ref{fig:momxj}, solutions of higher values of $j$ store their angular momenta mainly in the hair, hence the deformation factor drops with $j$ and hairy solutions never overcome the one from extremal Kerr. We note that for such high spinning parameters, when fixed together with $\omega_s$, the deformation decreases with $\kappa$. Unlike the horizon area,  though, as the solutions approach the clouds there is an intersection point where this behavior flips and the horizon becomes more oblate as the Gauss curvature increases.

\section{Conclusions}\label{sec:Conclusions}

In this paper we extended our previous works on solitonic vacuum of a certain class of TMST to show the existence of hairy black hole solutions, which were naturally expected. A set of five different theories, each prescribed by the value of the target space Gauss curvature, were analyzed by constructing thousands of solutions with varying $\omega_s$ and $r_H$ with a fixed winding number $m=1$ spanning the whole domain of existence on their parameter spaces. 

The boundaries of the domain of existence are qualitatively the same as those for Kerr BHs with scalar hair. To the accuracy we can currently reach, the scalar cloud solutions are indistinguishable in the five theories as the scalar field on a Kerr background is orders of magnitude smaller than the absolute value of the Gauss curvature. The other boundaries become however fairly distinct as $\kappa$ varies within the limits adopted in the paper and the differences are of course expected to increase for larger absolute values of the Gauss curvature. For fixed horizon radius, the compact support in $\omega_s$ decreases with increasing $\kappa$, while the maximum mass of the set of solutions becomes greater. In particular for $\kappa=-5$ it was not possible to access some parts of the domain, and specially understand where the solitonic curve approaches the extremal hairy black holes. For larger absolute values of $\kappa$, than the ones considered in the paper, the differences are of course expected to inclrease.

In all theories the Kerr bound is violated for a large set of hairy black holes, which comes as no surprise as pure solitons form one boundary of the domain of existence which can be continuously approached. Interestingly, as the Gauss curvature becomes smaller, greater is the minimum value a solution achieves for the spin parameter. In our most extreme case, $\kappa=-5$, we could not find solutions for which $j<0.8$, even for very large radii as $r_H=0.3$ for which the set of solutions has quite small compact support in $\omega_s$. Unlike other families of curves with fixed radius, for this specific $\kappa$, the solutions start and end at the clouds. Hairier solutions are found for greater values of $\kappa$, i.e. the maximum value of the mass ratio between hair and hole increase with this curvature. The mass and angular momentum do not uniquely determine the solution in any of these theories, and in the non-uniqueness region one finds that for some fixed pairs of them there exists solutions for which the horizon area is greater than the Kerr counterpart, and solutions for which it is smaller. 

An investigation on the stability of these objects is out of the scope of this paper, but a few words are worth of
note. As in the general relativistic case, the nature of these solutions lies on the superradiant instability that arises
from the excitation of some scalar modes. In this class of theories this scalar corresponds to an extra
gravitational degree of freedom and is universally minimally coupled to all source fields. 
Therefore, such excitation is to be granted by the surrounding material, since relevant astrophysical black holes are not expected to
dwell unescorted.

This paper is solely concerned on the existence of these solutions, and there is of course much more that can be explored even within one particular theory defined by the value of the Gauss curvature. Most importantly is the observational signatures these objects might have which could guide the search not only for hairy black holes, but also for particular features of these alternative theories. Although the next decades seem very promising for gravitational wave astronomy, it is believed that much higher precision is needed in order to measure any deviations caused by scalar matter, and there is still much to be done on the theoretical grounds of emissions in scalar-tensor theories, see e.g. \cite{PhysRevD.90.065019, Cardoso:2014uka}. Alternatively, there are other arenas where one could find relevant deviations from general relativistic black holes, such as the shadow it casts - we have shown that for a fixed ratio $j$ different objects are found, with horizon areas that can be larger or smaller than Kerr. Geodesic motion around these objects shall be considerably different than the ones around bald black holes, since the test particle could be immerse in the scalar matter and furthermore excite different modes of it. Even though high resolution observation of stars orbiting the vicinity of compact objects is not likely to happen in the near future, there can be measurable imprints on the accretion disks that would surround these black holes.

 \section*{Acknowledgements}
LC and DD acknowledge financial support via an Emmy Noether Research Group funded by the German Research Foundation (DFG) under grant no. DO 1771/1-1. DD is indebted to the Baden-Wurttemberg Stiftung for the financial support of this research project by the Eliteprogramme for Postdocs.  SY would like to thank the University of Tuebingen for the financial support.  SY acknowledges financial support by the Bulgarian NSF Grant KP-06-H28/7. Networking support by the COST Actions  CA16104 and CA16214 is also gratefully acknowledged.

\appendix

\section{Field Equations}\label{section:Appendix}
\label{fea}
Einstein field equations in diagonal form as we used are displayed below together with the KGE.

\begin{align}
\label{efe1}
r^2\partial^2_rF_0+\partial^2_\theta F_0 &= -\frac{1}{2}\left[\left(r\partial_rF_0\right)^2+\left(\partial_\theta F_0\right)^2\right]-\partial_rF_2\left(\frac{r^2\partial_rF_0}{2}+\frac{r_H}{\mathcal{N}}\right) -2r\frac{\partial_rF_0}{\mathcal{N}}
-\partial_\theta F_0\left(\cot\theta+\frac{\partial_\theta F_2}{2}\right) \nonumber \\
&+\frac{4\Omega^2\psi^2}{e^{2F_0}\mathcal{N}^2}\left(m\omega+\omega_sr\right)^2-4e^{2F_1}Vr^2+\frac{e^{2F_2-2F_0}\sin^2\theta}{\mathcal{N}^2}\left[\left(r\partial_r\omega\right)^2+\left(\partial_\theta \omega\right)^2+\omega^2-2r\omega\partial_r\omega\right] 
\end{align}

\begin{align}
\label{efe2}
r^2\partial^2_rF_1+\partial^2_\theta F_1 &=-r\partial_rF_1+\frac{r\partial_rF_0}{2}\left(r\partial_rF_2+2\right)+\partial_rF_2\frac{r_H}{\mathcal{N}}+\partial_\theta F_0\left(\cot\theta+\frac{\partial_\theta F_2}{2}\right) 
-2\Omega^2\left[\left(r\partial_r\psi\right)^2+\left(\partial_\theta\psi\right)^2\right] \nonumber \\
&+ 2e^{2F_1}\Omega^2\left[\frac{e^{-2F_2}m^2}{\sin^2\theta}-\frac{e^{-2F_0}}{\mathcal{N}^2}\left(m\omega+\omega_sr\right)^2\right] +
\frac{e^{2F_2-2F_0}\sin^2\theta}{2\mathcal{N}^2}\left[\left(r\partial_r\omega\right)^2+\left(\partial_\theta \omega\right)^2+\omega^2-2r\omega\partial_r\omega\right]+ 2\frac{r_H}{r-r_h}
\end{align}

\begin{align}
\label{efe3}
r^2\partial^2_rF_2+\partial^2_\theta F_2 &= -\frac{1}{2}\left[\left(r\partial_rF_2\right)^2+\left(\partial_\theta F_2\right)^2\right]-\frac{r\partial_rF_2}{2}\left(r\partial_rF_0+4+\frac{2}{\mathcal{N}}\right)
-\partial_\theta F_2\left(\frac{\partial_\theta F_0}{2}+2\cot\theta\right)-r\partial_rF_0-\cot\theta\partial_\theta F_0 \nonumber \\
&-\frac{4e^{2F_1}\Omega^2m^2\psi^2}{e^{2F_2}\sin^2\theta}-4e^{2F_1}r^2V
+\frac{e^{2F_2-2F_0}\sin^2\theta}{\mathcal{N}^2}\left[\left(r\partial_r\omega\right)^2+\left(\partial_\theta \omega\right)^2+\omega^2-2r\omega\partial_r\omega\right]- 2\frac{r_H}{r-r_h}
\end{align}

\begin{align}
\label{efe4}
r^2\partial^2_r\omega+\partial^2_\theta\omega &= \frac{r\partial_r\omega}{2}\left(r\partial_rF_0-3r\partial_rF_2-6+\frac{2}{\mathcal{N}}\right) + \frac{r\omega}{2}\left(3\partial_rF_2-\partial_rF_0\right)+\frac{1}{2}\left(\partial_\theta F_0-3\partial_\theta F_2 -6\cot\theta\right) \nonumber \\
&+ \frac{4e^{2F_1-2F_2}\Omega^2m}{\sin^2\theta}\left(m\omega+\omega_sr\right)+\omega\left(2-\frac{r_H}{r-r_H}\right)
\end{align}

\begin{align}
\label{eompsi}
r^2\partial^2_r\psi+\partial^2_\theta\psi &= -\frac{\Omega^2\kappa\psi}{4}\left[\left(r\partial_r\psi\right)^2+\left(\partial_\theta\psi\right)^2\right] +\frac{r\partial_r\psi}{2}\left(r\partial_rF_0+r\partial_rF_2+2+\frac{2}{\mathcal{N}}\right)
+\partial_\theta\psi\left(\frac{\partial_\theta F_0}{2}+\frac{\partial_\theta F_2}{2}+\cot\theta\right) \nonumber \\
&e^{2F_1}\psi\left(1-\frac{\Omega^2\kappa\psi^2}{4}\right)\left(\frac{e^{-2F_2}m^2}{\sin^2\theta}-\frac{e^{-2F_0}}{\mathcal{N}^2}\left(m\omega+\omega_sr\right)^2\right) + \frac{e^{2F_1}\mu^2r\psi}{\Omega^2}.
\end{align}

\bibliographystyle{ieeetr}
\bibliography{biblio}

\end{document}